\documentclass{article}

    \usepackage[final, nonatbib]{neurips_2019}

\usepackage{amsmath}
\usepackage{algorithm}
\usepackage[noend]{algpseudocode}
\usepackage[title]{appendix}

\usepackage[pdftex]{graphicx}
\usepackage[utf8]{inputenc} % allow utf-8 input
\usepackage[T1]{fontenc}    % use 8-bit T1 fonts
\usepackage{hyperref}       % hyperlinks
\usepackage{url}            % simple URL typesetting
\usepackage{booktabs}       % professional-quality tables
\usepackage{amsfonts}       % blackboard math symbols
\usepackage{nicefrac}       % compact symbols for 1/2, etc.
\usepackage{microtype}      % microtypography

\title{A crossover code for high-dimensional composition}

\author{%
  Rich Pang\\
  Computational Neuroscience Center\\
  University of Washington\\
  Seattle, WA 98195\\
  \texttt{rpang at uw dot edu} \\
}

\begin{document}

\maketitle

\begin{abstract}
We present a novel way to encode compositional information in high-dimensional (HD) vectors. Inspired by chromosomal crossover, random HD vectors are recursively interwoven, with a fraction of one vector's components masked out and replaced by those from another using a context-dependent mask. Unlike many HD computing schemes, "crossover" codes highly overlap with their base elements' and sub-structures' codes without sacrificing relational information, allowing fast element readout and decoding by greedy reconstruction. Crossover is mathematically tractable and has several properties desirable for robust, flexible representation.
\end{abstract}

\section{Introduction}

A common problem faced by intelligent systems is how to encode objects of variable complexity in fixed dimensions. Ideally, similar objects should have similar codes, new objects should not require rearranging existing codes, and there should be no limit on object complexity. Distributing codes across high-dimensional (HD) vectors offers a solution with potential robustness, flexibility, and generalization \cite{Hinton:1984, Mikolov:2013}; but how does one store compositional relations, like order or binding, in such a code, e.g. to distinguish AB vs BA or ((A,B), (C,D)) vs ((A,D), (C,B)), respectively? While trained systems like deep networks have some implicit capacity for this \cite{Bahdanau:2014, Luong:2015, Wu:2016}, it is unclear how these might handle complexity beyond that of their training data and challenging to quantify internal distributed representations \cite{Lipton:2016}. An alternative approach, known as HD computing (HDC), is to assign \textit{random} HD vector codes to a set of "base" elements of the data (e.g. symbols), then impose \textit{a priori} rules for composing these into richer structures (e.g. words) \cite{Plate:1995, Kanerva:2009, Gayler:2004}. HDC's advantages are (1) random HD vectors tend to have very low overlap, enabling large element codebooks, (2) there is no hard limit on data complexity, and (3) encoding is often analytically tractable. While most systems must of course undergo some amount of training, HDC provides a compelling blueprint for flexible representation.

An ideal code should allow fast readout of both a composition's base elements and their relations, but most HDC schemes support one or the other. Sums of random HD vectors, for instance, have high overlap (low distance under usual metrics) to their elements, enabling element readout via overlaps with the sum \cite{Bloom:1970, Anderson:1973, Plate:1994}; but cannot encode relations. HDC operations like circular convolution \cite{Plate:1995}, permutation \cite{Gayler:1998, Sahlgren:2008}, or matrix multiplication \cite{Gosmann:2019}, yield codes with cue-recoverable relations (cue A recovers B from ((A,B),(C,D))), but low overlap to the composition's base elements, precluding fast element readout without cues (but see \cite{Rachkovskij:2001}); this also conflicts with our desire that AB's code should resemble A's more than C's. Here we present a new HD composition operation inspired by chromosomal crossover, where two vectors $X$ and $X'$ are ordered or bound by masking out a fraction of $X$'s components and replacing them with those from $X'$; relational information is encoded in the context-dependent mask $\mu_:(X, X')$. "Crossover" codes thus highly overlap with their base elements, allowing fast recovery of both elements and relations, and decoding by greedy reconstruction. Further, they evenly distribute information and exhibit object-complexity-invariant statistics, ideal for robust intelligent systems. We describe crossover for sequences first, then generalize to binding and trees.

% Most HDC schemes use codes in which relations among elements can be easily recovered, but only if one provides certain elements as cues (e.g. cue A recovers B from ((A,B),(C,D))). Typical relation-encoding operations include circular convolution \cite{Plate:1995}, permutation \cite{Gayler:1998, Sahlgren:2008}, or matrix multiplication \cite{Gosmann:2019}, which create relational codes that little overlap their elements' codes (e.g. AB's code is distinct from A's and B's). This low overlap means elements cannot be directly recovered from the relation code (although see \cite{Rachkovskij:2001}), and it conflicts with our desire that AB should resemble A and B more than C. Here, we present a new way (to our knowledge) to encode compositional relations, where two HD vectors $X$ and $X'$ are ordered or bound by masking out a fraction of $X$'s components and replacing them with those from $X'$, analogous to chromosomal crossover, and with relational information encoded in the context-dependent mask $\mu(X, X')$. As a result, "crossover" composition codes highly overlap with the substructures and elements used to build them, allowing direct recovery of base elements and decoding by greedy reconstruction. Further, they distribute information evenly across components and have marginal statistics invariant to object complexity, suggesting utility for robust intelligent systems. We describe crossover for sequences first, then generalize to binding and trees.

\section{Coding scheme}

Let $X$ be a vector of $N$ components, each of which can take any of $Z$ states: $x_j \in G \equiv \{1,...,Z\}$; and let $d(X, X')$ be the Hamming distance from $X$ to $X'$. For a symbol sequence $Y \equiv Y_{1:L} \equiv (y_1, ..., y_L)$, where $y_t \in D \equiv \{1, ..., M\}$, a dictionary of $M$ symbols, we map $Y$ to $X^Y$ as follows:

We first sample a codebook $C \equiv \{X^1, ..., X^M, X^*\}$, where $X^*$ is a "start code", and $x_j$ are i.i.d. across all components and symbols and sampled uniformly from $G$. Next, we sample a mask function $\mu$, an $N \times Z^{2N}$ matrix with i.i.d. real-valued elements from $\textrm{Uniform}(0, 1)$, referenced as $\mu_j(X, X')$. In practice, columns of $\mu$ can be sampled "as needed" using a pseudorandom number generator.

Using $C$ and $\mu$ we encode sequences as shown in Algorithm 1 and Figure 1a: starting with $X^*$, we sequentially "weave" in symbol codes $X^{y_1}$, .., $X^{y_L}$ via an operation "CrossoverSeq": for the $t$'th element we mask out $1/(t+1)^\gamma$ of the evolving code's components and replace them with those from $X^{y_t}$. The mask selects components where $\mu_j(X^{Y_{1:t-1}}, X^{y_t}) < 1/(t+1)^\gamma$, so is unique for all $(X^{Y_{1:t-1}}, X^{y_t})$, yielding context-dependent mixing. $\gamma \geq 1$ prevents decay of early symbols in $X^Y$.

\begin{algorithm}
\caption{Sequence Encoding}
\label{alg:1}

\begin{algorithmic}[0]
\Function{CrossoverSeq}{$X, X'; t$} $\rightarrow X''$

\State $mask \gets (\mu_:(X, X') < 1/(t+1)^\gamma) \quad$  \# context-dependent mask with about $1/(t+1)^\gamma$ 1's
\State $X'' \gets X$
\State $X''[mask] \gets X'[mask] \quad$  \# replace $X$'s masked components with those from $X'$

\EndFunction

\Function{EncodeSequence}{$Y$} $\rightarrow X^Y$

\State $X^Y \gets X^*$
\For{$t \in 1, ..., |Y|$}

\State $X^Y \gets \textrm{CrossoverSeq}(X^Y, X^{y_t}, t)$

\EndFor

\EndFunction

\end{algorithmic}
\end{algorithm}

\begin{figure}
  \centering
  \includegraphics{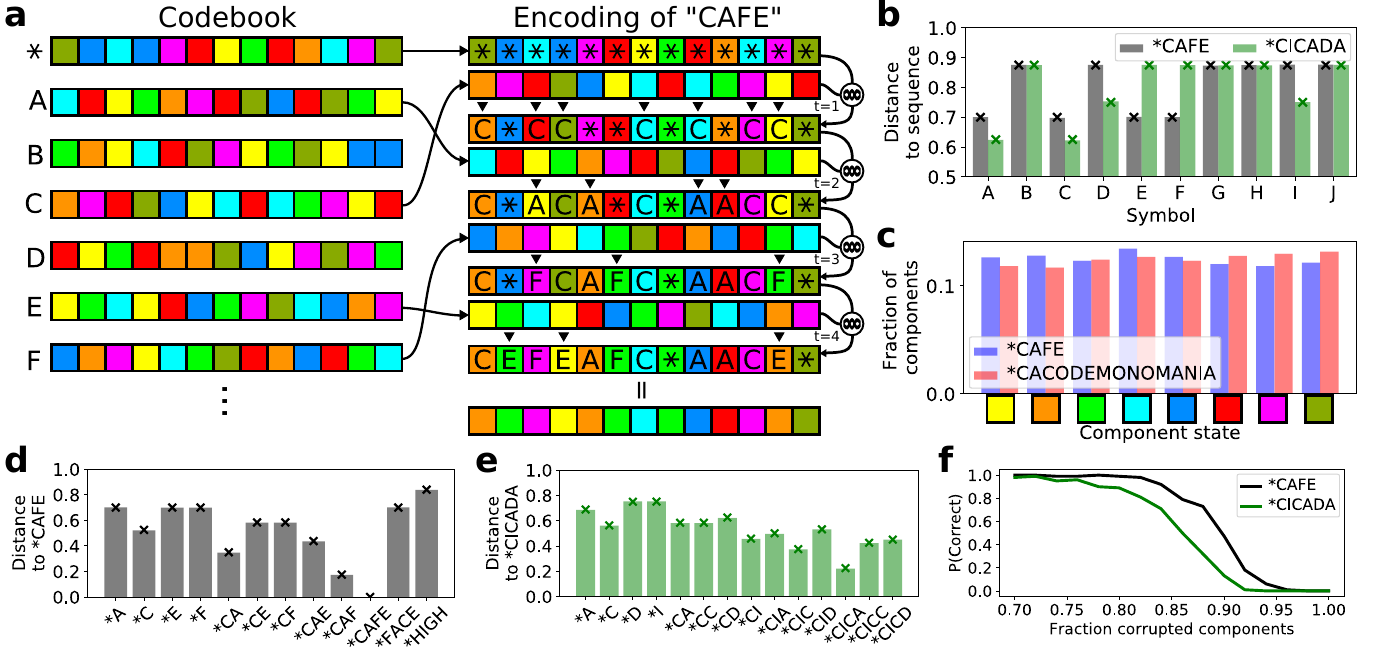}
  \caption{Sequence encoding via crossover ($N=1024, Z=8, M=26, \gamma=1$). (a) Example encoding of CAFE; colors are states; inscribed symbols are visual aids, not used by the algorithm. (b) Sequence-symbol code distances. (c) State distributions for example short and long sequence for one $C, \mu$ sample. (d, e) Distances of CAFE's and CICADA's code to other sequence codes. (f) Decoding vs code corruption. All plots here/elsewhere were made via 100 $C, \mu$ samples. "x"s denote exact calculations and bars average simulation results. "*" prefixes distinguish sequences from symbols.}
  \label{fig:1}
\end{figure}

For large $Z$ and $\gamma = 1$, mean similarity ($1-d$) of symbol code $X^i$ to the final sequence code $X^Y$ scales with $i$'s count within the sequence, $n^i$ (Figure 1b), with variance $\sim 1/N$. Precisely, $d(X^Y, X^i) \sim \textrm{Binomial}(N, 1-q)/N$, with $q \equiv P(x^Y_j = x^i_j) = n^i/(L+1) + (1/Z)(L+1-n^i)/(L+1)$. As exemplified by Figure 1c, crossover code component distributions are uniform and i.i.d. for a given $Y$, and invariant to sequence statistics (e.g. length); this results from the i.i.d. nature of $C, \mu$ and the interchangeability of component states, suggesting crossover as a robust, scale-invariant encoding.

For two sequences $Y$ and $Y'$ sharing their first $t$ symbols, $d(X^Y, X^{Y'})$ is also a scaled Binomial variable, and $q$ can be computed exactly for arbitrary $\gamma$; we present the $\gamma = 1$ case here. Let $\mathbf{n}_u$ be the length-$M$ symbol count vector for $Y_{1:t} = Y'_{1:t}$ with $\mathbf{1}^T\mathbf{n}_u = t$, and $\mathbf{n}_v$ and $\mathbf{n}_{v'}$ be the symbol count vectors for $Y_{t+1:L}$ and $Y'_{t+1:L'}$, respectively, with $\mathbf{1}^T\mathbf{n}_{v} = L - t$ and $\mathbf{1}^T\mathbf{n}_{v'} = L' - t$. Then:
$$q \equiv P(x_j^Y = x_j^{Y'}) = \frac{t+1}{L+1}\frac{t+1}{L'+1} + \frac{1}{L+1}\frac{1}{L'+1}\left[\mathbf{n}^T_v\mathbf{n}_{v'} + \left((L-t)(L'-t) - \mathbf{n}^T_v\mathbf{n}_{v'} \right)\frac{1}{Z}\right]$$
$$+\frac{1}{L+1}\frac{1}{L'+1}\left[
\mathbf{n}^T_{u}(\mathbf{n}_{v} + \mathbf{n}_{v'}) + \left((t+1)(L'+L-2t) - \mathbf{n}^T_{u}(\mathbf{n}_{v} + \mathbf{n}_{v'})
 \right)\frac{1}{Z}
\right]$$
For most $Y$, $X^Y$ overlaps little with codes for $Y$'s anagrams, and more with correct than incorrect starting subsequences, e.g. CI's and CIC's codes are closer than CD's and CIA's codes, respectively, to CICADA's code (Figure 1d,e). This is because different next-symbols yield different $\mu$, which quickly decreases overlap with $X^Y$ for sequence codes with out-of-order symbols. Consequently, $Y$ can be "greedily" decoded by rebuilding $X^Y$ from $X^*$, outputting at the $t$-th rebuild step the symbol $i$ that maximizes overlap with $X^Y$ when $X^i$ is woven into the current code, then weaving in $X^i$ and advancing to $t+1$. Decoding is robust to noise added to $X^Y$, and when a prior $P(Y)$ exists, the mutual information between $x_j^Y$ and $Y$ is i.i.d. across $j$; robustness is thus equivalent for random and targeted attacks, with decoding accuracy depending only on the number of corrupted $x_j^Y$ (Figure 1f).

While crossover intrinsically supports sequences with repeated symbols (either consecutive or not), for $Y$ with many symbol repeats distinct from $y_1$ (e.g. REFEREE), $X^*$ crossed over with $X^{y_1}$ can be further from $X^Y$ than $X^*$ crossed with the repeated symbol, impairing decoding. Making $\gamma > 1$ fixes this by more heavily weighting early symbols (Figure 2a). Surprisingly, while this front-loaded weighting improves decoding of, say, ABBBBBB, it does not equivalently degrade decoding of AAAAAAB, and optimal $\gamma$ are shared by many sequences (Figure 2b). In general, decoding accuracy decreases (increases) with $L$ ($N$), since $\textrm{Var}[P(x_j^{Y_{1:t}} = x_j^Y)] \sim 1/N$. Decoding accuracy is also robust to large $M$, the dictionary size, due at heart to the low expected overlap of HD random vectors.

\begin{figure}
\label{fig:2}
  \centering
  \includegraphics{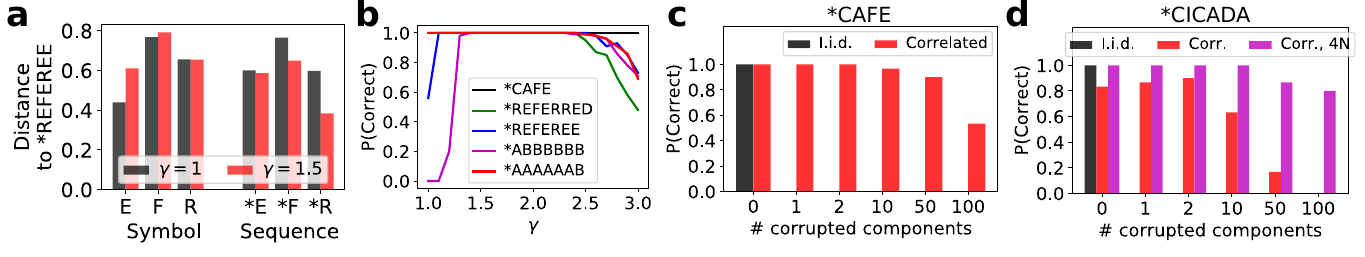}
  \caption{Crossover modifications ($N=1024, Z=8, M=26$). (a) Distance of *REFEREE to symbols/sequences for two $\gamma$. (b) Decoding vs. $\gamma$ for several sequences. (c) Decoding of *CAFE vs number of corrupted components at each encoding step for i.i.d. vs correlated $\mu$. (d) Decoding of *CICADA for i.i.d. $\mu$, correlated $\mu$, and correlated $\mu$ with quadrupled $N$.}
\end{figure}

While crossover with i.i.d. $\mu$ is robust to noise added to the final code (Figure 1f), it fails when noise is added during encoding (Figure 2c,d), as one altered component yields an entirely different mask. This is fixed by introducing correlations into $\mu_j(X, X')$ (i.e. making $\mu$ a smooth function of $X, X'$ [see Appendix C]). The cost of this is decreased decoding accuracy for certain sequences, since correlated are less distinct than i.i.d. masks, but this can be corrected by increasing $N$ (Figure 2c,d).

Beyond sequences, crossover naturally generalizes to \textit{k}-ary trees. Key to encoding trees is \textit{binding} elements into groups that can themselves be bound, yet without forgetting initial group identities: ((A,B), (C,D)) must be encoded differently from ((A,D), (C,B)). In crossover, $k$ elements $i_1, ..., i_k$ can be bound into a group using a context-dependent mask function $\mu_j(X^{i_1}, ..., X^{i_k})$ (which can optionally commute). The vector for the group takes components from $X^{i_1}$ where $0 < \mu < 1/k$, from $X^{i_2}$ where $1/k < \mu < 2/k$, etc. This can then be bound itself in the exact same manner (Figure 3a), allowing recursive encoding of an arbitrary tree. The result is a vector overlapping more with elements in vs not in the tree, and with groups in vs not in the tree, even if the latter contain the same base elements (Figure 3b). A generic tree encoding will also tend to overlap more with its included sub-trees than with alternative sub-trees of equal complexity. Similar to sequence encoding, this means both base elements and their recursive relations can be extracted via their overlap with the final code, and that decoding of the full tree can occur through recursive but greedy reconstruction.

\begin{figure}
\label{fig:3}
  \centering
  \includegraphics{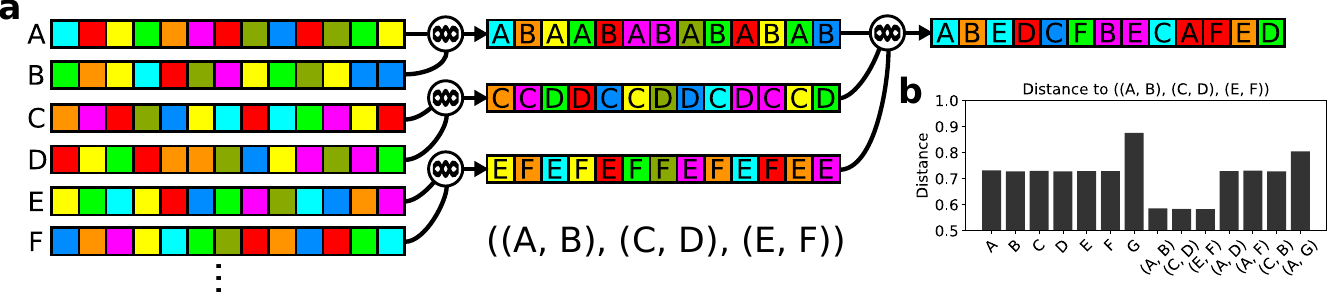}
  \caption{Tree/binding encoding via crossover ($N=1024, Z=8, M=26$). (a) Example crossover encoding of ((A,B), (C,D), (E,F)). (b) Distances of final code to element and pair/group codes.}
\end{figure}
\section{Discussion}

While intelligent systems must typically undergo at least some amount of training to function, it can be useful for certain computational features to be built in \cite{Zador:2019}. Flexible compositional coding is of particular relevance as it provides a substrate for computing with novel objects or events if they are made from familiar elements: an autonomous vehicle should be able to manipulate a representation of "child on scooter behind car in front of bus", even if previously it had only worked with the isolated scene elements. While many trained systems, e.g. translation networks \cite{Bahdanau:2014, Luong:2015, Wu:2016}, have some implicit capacity for compositional encoding (otherwise outputs would be disordered), it is unclear how they could handle complexity beyond their training data, and understanding their internal representations can be challenging \cite{Lipton:2016}, although exciting progress is being made in interpretability \cite{Zeiler:2014, Montavon:2018}. Alternatively, one can design HDC algorithms with \textit{a priori} flexibility, then attach or embed them in trainable systems. Basing training on convolutional binding, for instance, allowed an artificial network to efficiently and scalably learn knowledge graphs and out-perform state-of-the-art methods at link prediction \cite{Nickel:2016}. It stands to reason that identifying robust HDC codes could prove useful to improving artificial systems, and potentially guiding investigation of neural computation in biology.

"Crossover" shares advantages with existing HDC schemes but is distinct in key ways also. As in other schemes \cite{Kanerva:2009}, crossover preserves recoverable relational information, and relation codes can be reused as elements, allowing arbitrarily deep composition. For instance, in crossover one can add CAFE's or other words' codes to the dictionary (perhaps with minor noise added), enabling encoding of long sequences made from short ones. To encode relations, previous schemes have used circular convolution \cite{Plate:1995}, XOR (for binary vectors) \cite{Kanerva:1994}, permutation \cite{Sahlgren:2008, Gayler:1998}, or matrix multiplication \cite{Gosmann:2019}, which yield relation codes that little overlap their elements. While these distinguish different compositions of the same elements and allow cue-based recall, base elements cannot be read out directly from their overlap with the relation code, and the entire dictionary must be used (instead of just the relevant symbols) when reading out relations, potentially greatly slowing decoding for large $M$. An exception is "context-dependent thinning" (CDT), which stores relations via the union of the elements, then "stamps" the code with a context-dependent set of zeros \cite{Rachkovskij:2001}; while the stamp shape retains relational information without erasing element overlaps, however, information beneath its imprint is discarded, and it is not clear how CDT could construct recursive sequence codes without early elements decaying exponentially. Crossover overcomes these limitations, as both elements and their relations can be read out directly, and in sequences no element overlaps decay exponentially. Further, crossover yields i.i.d. information across code components, and code statistics are invariant to object complexity, suggesting it may be a useful representation for a robust intelligent system.

Crossover is also conceptually distinct from its predecessors since one need not define arithmetical operations over code components. Whereas previous HDC algorithms require \textit{summing} components, either directly or through convolution \cite{Plate:1995, Gayler:1998, Sahlgren:2008, Gosmann:2019, Rachkovskij:2001, Kanerva:1994}, crossover's component states $1, ..., Z$ are just labels, with no numeric value; they need only a comparison operation. Crossover thus outlines a distributed computation substrate that can exist outside vector spaces, and instead within more generic metric spaces. This is relevant to biological implementation, as it allows one to envision $x_j$ not only as firing rates of neurons or populations, but also as neural assembly states that might not make sense to add, multiply, or rank. For instance, each component of a crossover code could correspond to the attractor state of a small Hopfield-like assembly \cite{Hopfield:1982}, with the collective state of a pool of $N$ such assemblies encoding a complex data object. This could serve, e.g., as a crude implementation of a combinatorially expressive working memory substrate, yet with sufficient code-overlap statistics to be usefully read out and manipulated by further downstream and recurrent computations.

\subsubsection*{Acknowledgments}

I would like to thank Taliesin Beynon for many helpful suggestions in contextualizing this work within the broader scope of high-dimensional computing and vector symbolic architectures.

% Use unnumbered third level headings for the acknowledgments. All acknowledgments
% go at the end of the paper. Do not include acknowledgments in the anonymized
% submission, only in the final paper.

\bibliographystyle{unsrt}  
\bibliography{references}  %%% Remove comment to use the external .bib file (using bibtex).

\begin{thebibliography}{10}

\bibitem{Hinton:1984}
Geoffrey~E Hinton, James~L McClelland, David~E Rumelhart, et~al.
\newblock {\em Distributed representations}.
\newblock Carnegie-Mellon University Pittsburgh, PA, 1984.

\bibitem{Mikolov:2013}
Tomas Mikolov, Ilya Sutskever, Kai Chen, Greg~S Corrado, and Jeff Dean.
\newblock Distributed representations of words and phrases and their
  compositionality.
\newblock In {\em Advances in neural information processing systems}, pages
  3111--3119, 2013.

\bibitem{Bahdanau:2014}
Dzmitry Bahdanau, Kyunghyun Cho, and Yoshua Bengio.
\newblock Neural machine translation by jointly learning to align and
  translate.
\newblock {\em arXiv preprint arXiv:1409.0473}, 2014.

\bibitem{Luong:2015}
Minh-Thang Luong, Hieu Pham, and Christopher~D Manning.
\newblock Effective approaches to attention-based neural machine translation.
\newblock {\em arXiv preprint arXiv:1508.04025}, 2015.

\bibitem{Wu:2016}
Yonghui Wu, Mike Schuster, Zhifeng Chen, Quoc~V Le, Mohammad Norouzi, Wolfgang
  Macherey, Maxim Krikun, Yuan Cao, Qin Gao, Klaus Macherey, et~al.
\newblock Google's neural machine translation system: Bridging the gap between
  human and machine translation.
\newblock {\em arXiv preprint arXiv:1609.08144}, 2016.

\bibitem{Lipton:2016}
Zachary~C Lipton.
\newblock The mythos of model interpretability.
\newblock {\em arXiv preprint arXiv:1606.03490}, 2016.

\bibitem{Plate:1995}
Tony~A Plate.
\newblock Holographic reduced representations.
\newblock {\em IEEE Transactions on Neural networks}, 6(3):623--641, 1995.

\bibitem{Kanerva:2009}
Pentti Kanerva.
\newblock Hyperdimensional computing: An introduction to computing in
  distributed representation with high-dimensional random vectors.
\newblock {\em Cognitive computation}, 1(2):139--159, 2009.

\bibitem{Gayler:2004}
Ross~W Gayler.
\newblock Vector symbolic architectures answer jackendoff's challenges for
  cognitive neuroscience.
\newblock {\em arXiv preprint cs/0412059}, 2004.

\bibitem{Bloom:1970}
Burton~H Bloom.
\newblock Space/time trade-offs in hash coding with allowable errors.
\newblock {\em Communications of the ACM}, 13(7):422--426, 1970.

\bibitem{Anderson:1973}
James~A Anderson.
\newblock A theory for the recognition of items from short memorized lists.
\newblock {\em Psychological Review}, 80(6):417, 1973.

\bibitem{Plate:1994}
Tony~A Plate.
\newblock {\em Distributed representations and nested compositional structure}.
\newblock University of Toronto, Department of Computer Science, 1994.

\bibitem{Gayler:1998}
Ross~W Gayler.
\newblock Multiplicative binding, representation operators \& analogy (workshop
  poster).
\newblock 1998.

\bibitem{Sahlgren:2008}
Magnus Sahlgren, Anders Holst, and Pentti Kanerva.
\newblock Permutations as a means to encode order in word space.
\newblock 2008.

\bibitem{Gosmann:2019}
Jan Gosmann and Chris Eliasmith.
\newblock Vector-derived transformation binding: An improved binding operation
  for deep symbol-like processing in neural networks.
\newblock {\em Neural computation}, 31(5):849--869, 2019.

\bibitem{Rachkovskij:2001}
Dmitri~A Rachkovskij and Ernst~M Kussul.
\newblock Binding and normalization of binary sparse distributed
  representations by context-dependent thinning.
\newblock {\em Neural Computation}, 13(2):411--452, 2001.

\bibitem{Zador:2019}
Anthony~M Zador.
\newblock A critique of pure learning and what artificial neural networks can
  learn from animal brains.
\newblock {\em Nature communications}, 10(1):1--7, 2019.

\bibitem{Zeiler:2014}
Matthew~D Zeiler and Rob Fergus.
\newblock Visualizing and understanding convolutional networks.
\newblock In {\em European conference on computer vision}, pages 818--833.
  Springer, 2014.

\bibitem{Montavon:2018}
Gr{\'e}goire Montavon, Wojciech Samek, and Klaus-Robert M{\"u}ller.
\newblock Methods for interpreting and understanding deep neural networks.
\newblock {\em Digital Signal Processing}, 73:1--15, 2018.

\bibitem{Nickel:2016}
Maximilian Nickel, Lorenzo Rosasco, and Tomaso Poggio.
\newblock Holographic embeddings of knowledge graphs.
\newblock In {\em Thirtieth Aaai conference on artificial intelligence}, 2016.

\bibitem{Kanerva:1994}
Pentti Kanerva.
\newblock The spatter code for encoding concepts at many levels.
\newblock In {\em International Conference on Artificial Neural Networks},
  pages 226--229. Springer, 1994.

\bibitem{Hopfield:1982}
John~J Hopfield.
\newblock Neural networks and physical systems with emergent collective
  computational abilities.
\newblock {\em Proceedings of the national academy of sciences},
  79(8):2554--2558, 1982.

\end{thebibliography}

\begin{appendices}
\section{Properties of crossover codes}

\subsection{Proof that sequence codes have i.i.d. components}

Denoting $Y_{1:t} \equiv (y_1, ..., y_t)$ with $Y_{1:L} = Y$, and $C_j \equiv (x^1_j, ..., x^M_j, x^*_j)$, we show that $x_j^{Y_{1:t}}$ are independent across $j$ for all $t$.

We will use the fact that if $A_1, ..., A_N$ are independent, then so are $f(A_1), ..., f(A_N)$. In particular, define $\xi^t_j \equiv \{x_j^{Y_{1:t-1}}, ..., x_j^{Y_{1:0}}, C_j, \mu(j, X^*, X^{y_1}), \mu(j, X^{Y_{1:1}}, X^{y_2}), ..., \mu(j, X^{Y_{1:t-1}}, X^{y_t})\}$. Then $x_j^{Y_{1:t}}$ is a strict function of $\xi^t_j$ so it suffices to show $\xi^t_j$ are independent across $j$ for all $t$.

We proceed by induction:

\textbf{Base case:} $\xi^1_j = \{x^{Y_{1:0}}_j, C_j, \mu(j, X^{Y_{1:0}}, X^{y_1})\} = \{x^*_j, C_j, \mu(j, X^*, X^{y_1})\}$, which are all i.i.d., so $\xi^1_j$ are independent across $j$.

\textbf{Inductive step:} If $\xi^t_j$ are independent across $j$ then so are $\{x_j^{Y_{1:t}}\} \cup \xi^t_j$, since this is a deterministic function of $\xi^t_j$. Further, $\{x_j^{Y_{1:t}}\} \cup \xi^t_j \cup \{\mu(j, X^{Y_{1:t}}, X^{y_{t+1}})\}$ must also be independent across $j$. To see this, consider the following two cases.

First, most likely is that $\mu(j, X^{Y_{1:t}}, X^{y_{t+1}})$ has never been used in the construction of $X^{Y_{1:t-1}}$, so it is independent from $\{x_j^{Y_{1:t}}\} \cup \xi^t_j$ while also independent across $j$. Therefore, their union is independent across $j$.

In the second, less likely case, $\mu(j, X^{Y_{1:t}}, X^{y_{t+1}})$ \textit{was} used to construct $X^{Y_{1:t-1}}$ at some point in the past, i.e. $(X^{Y_{1:t}}, X^{y_{t+1}}) = (X^{Y_{1:t'}}, X^{y_{t'+1}})$ for some $t' < t$. In this case, however, the term $\mu(j, X^{Y_{1:t}}, X^{y_{t+1}}) = \mu(j, X^{Y_{1:t'}}, X^{y_{t'+1}})$ is already included in $\xi^t_j$, and so the independence across $j$ is not affected because no new random variables are added. Thus, $\{x_j^{Y_{1:t}}\} \cup \xi^t_j \cup \{\mu(j, X^{Y_{1:t}}, X^{y_{t+1}})\}$ is independent across $j$.

Since $\{x_j^{Y_{1:t}}\} \cup \xi^t_j \cup \{\mu(j, X^{Y_{1:t}}, X^{y_{t+1}})\} = \xi^{t+1}_j$, we have that $\xi^t_j$ being independent across $j$ implies that $\xi^{t+1}_j$ is independent across $j$, completing the inductive step.

Therefore $x_j^{Y_{1:t}}$ is independent across $j$ for all $t$, in other words $P(X^Y|Y) = \prod\limits_{j=1}^N P(x^Y_j|Y)$.

The equality $P(x^Y_j|Y) = P(x^Y_{j'}|Y) \quad \forall j, j'$ arises because no component is treated differently from any other in our algorithm, so symmetry implies they can have no statistical differences.

\subsection{Proof that component distributions are uniform over component states}

$P(x^Y_j = g|Y) = P(x^Y_j = g'|Y)$ for $g \neq g'$ arises because all $g$ in our model are treated identically, so symmetry implies any functions of them can have no statistical differences. Since these must add to $1$ when summed over $g \in G$, and $|G| = Z$, we must also have $P(x^Y_j = g|Y) = 1/Z.$

This also implies that the marginal statistics of crossover codes do not depend on $Y$ or any feature of $Y$ e.g. sequence length. This distinguishes crossover, for instance, from representations that would grow denser or sparser with the length of the sequence they represent.

\subsection{Proof that sequence information is i.i.d. across components}

Given $C$, $\mu$, and a prior distribution over sequences $P_Y(Y)$ with entropy $H[Y]$ over $N_Y$ different sequences in total, define the information $I_j | C, \mu$ in component $j$ as the mutual information between $Y$ and $x_j$, i.e. 

$$I_j | C, \mu \equiv MI[Y|x_j; C, \mu] = H[Y] - H[Y|x_j; C, \mu] = $$
$$H[Y] - \sum_g H[Y|x_j = g; C, \mu] P_Y(x_j = g| C, \mu).$$

The components for one sequence $Y$ are i.i.d. Therefore, while the $x^Y_j$ and $x^{Y'}_j$ might be dependent (e.g. if $Y$ and $Y'$ are similar), there can be no dependencies between $x^Y_j$ and $x^{Y'}_{j'}$ for $j \neq j'$. Therefore $\{x^{Y_1}_j, ..., x^{Y_{N_Y}}_j\}$ is independent across $j$. Since $I_j$ is a deterministic function of $\{x^{Y_1}_j, ..., x^{Y_{N_Y}}_j\}$ (quantifying how many possible sequences get eliminated upon knowing the value of $x^Y_j$), $I_j$ is independent across $j$ also. Since no $j$ is treated specially in our algorithm, $I_j$ must also be identically distributed, so given an initial prior over sequences to encode $P_Y(Y)$, $I_j$ is i.i.d.

\section{Code overlap derivations for symbols and sequences}

Since $x^Y_j$ are i.i.d., the similarity, or overlap, between two sequence codes  (1 minus the Hamming distance) $s(X^Y, X^{Y'}) \sim \frac{1}{N}\textrm{Bi}(N, q)$, where $\textrm{Bi}(N, q)$ is a binomial distribution with success probability $q = P(x^Y_j = x^{Y'}_j)$. The randomness giving rise to the distribution is over samples of $C$ and $\mu$. Thus, $\textrm{E}[s(X^Y, X^{Y'})] = q$ and $\textrm{Var}[s(X^Y, X^{Y'})] = q(1-q)/N$, and it suffices to only compute $q$. Similar results hold for the similarity of a sequence and symbol code, $s(X^Y, X^i)$. For conciseness, we drop the $j$ and write $x \equiv x_j$ from now on.

Our goal is to compute $q$ for an arbitrary sequence $Y$ and symbol $i$ and for two arbitrary sequences $Y$ and $Y'$, since this gives us the full distribution of similarity values over samples of $C$ and $\mu$.

\subsection{Useful quantities}

We first compute a few quantities to keep calculations more concise.

1. $f^{\gamma,L}_t \equiv P(x^Y \leftarrow y_t)$ is the probability that $x^Y$ was taken from the $t$-th element of $Y$. This is $1/(t+1)^\gamma$ times the probability the component was not overwritten between $t+1$ and $L$ (including occasions where it was overwritten by the same state). We then have:

$$f^{\gamma,L}_t \equiv P(x^Y \leftarrow y_t) = \frac{1}{(t+1)^\gamma}\prod\limits_{t' = t+1}^L\left(1 - \frac{1}{(t'+1)^\gamma}\right).$$

Note that this is not a function of $x^Y$. Also note that $f^{1,L}_t = 1/(L+1)$, which is independent of $t$.

2. $P(x^Y \leftarrow i)$ is the probability that $x^Y$ was taken from symbol $i$, which will depend on how many times $i$ was present in $Y$. We compute this by summing the probabilities of all ways this could have occurred.

Suppose $y_t = i$ for $t_1 < ... < t_{n^i}$, with $\sum_i n^i = L$. We can divide the combinatorial ways $x^Y$ could have been taken from $i$ into $n^i$ cases. First, $x^Y$ could have been taken from $y_{t_1}$ then never overwritten, whose probability is $f^{\gamma,L}_{t_1}$. Next, it could have been taken from $y_{t_2}$ then never overwritten (regardless of whether it was taken from $y_{t_1}$), whose probability is $f^{\gamma,L}_{t_2}$. And so forth until $t_{n^i}$. These $n^i$ cases are mutually non-overlapping and cover all possible ways that $x^Y$ could have been taken from $i$. Thus

$$P(x^Y \leftarrow i) = f^{\gamma,L}_{t_1} + ... + f^{\gamma,L}_{t_{n^i}}$$.

When $\gamma = 1$, $P(x^Y \leftarrow i) = n^i/(L+1)$.

3. $P(x^Y \leftarrow Y_{1:t})$ is the probability that $x^Y$ was not overwritten from $t+1$ to $L$. This is just

$$P(x^Y \leftarrow Y_{1:t}) = \prod\limits_{t' = t+1}^L \left(1 - \frac{1}{(t'+1)^\gamma}\right) = (t+1)^\gamma f^{\gamma,L}_t.$$

When $\gamma = 1$, $P(x^Y \leftarrow Y_{1:t}) = (t+1)/(L+1)$.

4. $P(x^Y \leftarrow i|x^Y \not \leftarrow Y_{1:t})$ is the probability that $x^Y$ came from $i$ given that it did not come from $Y_{1:t}$ (i.e. given that it was overwritten at some point from $t+1$ to $L$). Suppose $y_s = i$ for $t < s_1 < ... < s_{n^i_v}$ where $n^i_v$ is the number of times symbol $i$ appears in $Y_{t+1:L} \equiv (y_{t+1}, ..., y_L)$, with $\sum\limits_i n^i_v = L-t$.

We compute $P(x^Y \leftarrow i|x^Y \not \leftarrow Y_{1:t})$ by summing over the probabilities of all the ways this could have occurred. Similar to the calculation for $P(x^Y \leftarrow i)$, we have

$$P(x^Y \leftarrow i|x^Y \not \leftarrow Y_{1:t}) = P(x^Y \leftarrow y_{s_1}| x^Y \not \leftarrow Y_{1:t}) + ... + P(x^Y \leftarrow y_{s_{n^i_v}}| x^Y \not \leftarrow Y_{1:t}).$$

By Bayes' Rule:

$$P(x^Y \leftarrow y_{s_k}| x^Y \not \leftarrow Y_{1:t}) = \frac{P(x^Y \not \leftarrow Y_{1:t}| x^Y \leftarrow y_{s_k})P(x^Y \leftarrow y_{s_k})}{P(x^Y \not\leftarrow Y_{1:t})} = \frac{P(x^Y \leftarrow y_{s_k})}{P(x^Y \not\leftarrow Y_{1:t})}$$

since if $x^Y$ came from $y_{s_k}$, which is in the part of $Y$ after $t$, then $x^Y$ could have not have come from $Y_{1:t}$. We have already computed the numerator and denominator (which is just $1 - P(x^Y \leftarrow Y_{1:t}))$, so

$$P(x^Y \leftarrow i|x^Y \not \leftarrow Y_{1:t}) = \frac{1}{P(x^Y \not\leftarrow Y_{1:t})}\left( f^{\gamma,L}_{s_1} + ... + f^{\gamma,L}_{s_{n^i_v}} \right).$$

When $\gamma = 1$ this reduces to $P(x^Y \leftarrow i|x^Y \not \leftarrow Y_{1:t}) = n^i_v/(L-t)$, i.e. it is proportional to how many times $i$ appears in the second part of the sequence $Y_{t+1:L}$.

We now calculate how similar a sequence code is to any symbol code, as well as to other sequence codes.

\subsection{Similarity between symbol and sequence codes}

$$P(x^Y = x^i) = P(x^Y \leftarrow i)P(x^Y = x^i|x^Y \leftarrow i) + P(x^Y \not \leftarrow i)P(x^Y = x^i|x^Y \not \leftarrow i)
$$
$$= P(x^Y \leftarrow i) + P(x^Y \not \leftarrow i)\frac{1}{Z}$$

$$= \sum\limits_{t \in (t_1, ..., t_{n^i})}f^{\gamma,L}_t + \left(1 - \sum\limits_{t \in (t_1, ..., t_{n^i})}f^{\gamma,L}_t\right)\frac{1}{Z}
= \sum\limits_{t \in (t_1, ..., t_{n^i})}\left(1 - \frac{1}{Z}\right)f^{\gamma,L}_t + \frac{1}{Z}$$

When $\gamma = 1$, $f_t^{1,L} = 1/(L+1)$ so $P(x^Y = x_i)$ simplifies to $n^i/(L+1) + (1/Z)(L+1-n^i)/(L+1)$, which tends to  $n^i/(L+1)$ for large $Z$.

\subsection{Similarity between codes for two sequences}

Let $u^i$ index the times $i$ appears in $Y_{1:t}$, $v^i$ the times $i$ appears in $Y_{t+1:L}$, and $(v')^i$ the times $i$ appears in $Y'_{t+1:L'}$. Let $n^i_u$ count how many times $i$ appears in $Y_{1:t}$, $n^i_v$ the number of times $i$ appears in $Y_{t+1:L}$, and $n^i_{v'}$ the number of times $i$ appears in $Y'_{t+1:L'}$. We also write these in vector format as $\mathbf{n}_u$, $\mathbf{n}_v$, and $\mathbf{n}_{v'}$, respectively, where $\mathbf{n}_u$, $\mathbf{n}_v$, and $\mathbf{n}_{v'}$ are length-$M$ vectors, with one component per symbol in the dictionary $D$.

Since $Y_{1:t} = Y'_{1:t}$,  first note that $X^{Y_{1:t}} = X^{Y'_{1:t}}$, so $x^{Y_{1:t}} = x^{Y'_{1:t}}$. We wish to find $P(x^Y = x^{Y'})$.

The event $x^Y = x^{Y'}$ can occur in 4 distinct ways, whose probabilities sum to $P(x^Y = x^{Y'})$. At present we assume a negligible chance two sequence or symbol codes are exactly identical.

\textbf{Case 1}: $(x^Y \leftarrow Y_{1:t}) \land (x^{Y'} \leftarrow Y'_{1:t})$, i.e. neither $x$ is overwritten at $t+1$ or later. These events are independent, so

$$P(\textrm{Case 1}) = P(x^Y \leftarrow Y_{1:t})P(x^{Y'} \leftarrow Y'_{1:t}) = (t+1)^\gamma f^{\gamma,L}_t(t+1)^\gamma f^{\gamma,L'}_t$$

When $\gamma = 1$, then:

$$P(\textrm{Case 1}) = \frac{t+1}{L+1}\frac{t+1}{L'+1}$$.

\textbf{Case 2}: $(x^Y \not\leftarrow Y_{1:t}) \land (x^{Y'} \not\leftarrow Y'_{1:t}) \land (x^Y = x^{Y'})$, i.e. $x$ is overwritten in both sequences at some point between $t+1$ and $L$ or $L'$, but by the same component state. Since the first two conditions are again independent we have

$$P(\textrm{Case 2}) = P(x^Y \not\leftarrow Y_{1:t})P(x^{Y'} \not\leftarrow Y'_{1:t})P(x^Y = x^{Y'}|x^Y, x^{Y'} \not\leftarrow Y_{1:t})$$

$$= [1 - (t+1)^\gamma f^{\gamma,L}_t][1 - (t+1)^\gamma f^{\gamma,L'}_t]P(x^Y = x^{Y'}|x^Y, x^{Y'} \not\leftarrow Y_{1:t})$$

where we recall that $Y_{1:t} = Y'_{1:t}$. When $\gamma = 1$ this simplifies to

$$P(\textrm{Case 2}) = \frac{L-t}{L+1}\frac{L'-t}{L'+1}P(x^Y = x^{Y'}|x^Y, x^{Y'} \not\leftarrow Y_{1:t})$$

To find $P(x^Y = x^{Y'}|x^Y, x^{Y'} \not\leftarrow Y_{1:t})$ we note that if both $x$'s are overwritten, they can end up with the same component state if (1) $x^Y$ and $x^{Y'}$ are taken from the same symbol $i$, whose probability we write as $P(x^Y, x^{Y'} \leftarrow same|x^Y, x^{Y'} \not\leftarrow Y_{1:t})$ or (2) $x^Y$ and $x^{Y'}$ are taken from different symbols but end up with the same component state by chance. That is,

$$P(x^Y = x^{Y'}|x^Y, x^{Y'} \not\leftarrow X^{Y_{1:t}}) =$$

$$P(x^Y, x^{Y'} \leftarrow same|x^Y, x^{Y'} \not\leftarrow X^{Y_{1:t}}) + [1 - P(x^Y, x^{Y'} \leftarrow same|x^Y, x^{Y'} \not\leftarrow Y_{1:t})]\frac{1}{Z}.$$

$P(x^Y, x^{Y'} \leftarrow same|x^Y, x^{Y'} \not\leftarrow X^{Y_{1:t}})$ is the sum over $i \in \{1, ..., M\}$ of the probabilities that both $x^Y$ and $x^{Y'}$ were taken from $i$. These events are independent, so:

$$P(x^Y, x^{Y'} \leftarrow i|x^Y, x^{Y'} \not\leftarrow X^{Y_{1:t}}) = P(x^Y \leftarrow i|x^Y, x^{Y'} \not\leftarrow X^{Y_{1:t}})P(x^{Y'} \leftarrow i|x^Y, x^{Y'} \not\leftarrow X^{Y_{1:t}})$$

$$ = \frac{\sum\limits_{v^i} f_{v^i}^{\gamma,L}}{1 - (t+1)^\gamma f^{\gamma,L}_t} \frac{\sum\limits_{(v')^i} f^{\gamma,L'}_{(v')^i}}{1 - (t+1)^\gamma f^{\gamma,L'}_t}.$$

When $\gamma = 1$, this simplifies to

$$\frac{n^i_v}{L-t}\frac{n^i_{v'}}{L'-t}$$.

Thus

$$P(x^Y, x^{Y'} \leftarrow same|x^Y, x^{Y'} \not\leftarrow X^{Y_{1:t}}) = \sum_i \frac{\sum\limits_{v^i} f^{\gamma,L}_{v^i}}{1 - (t+1)^\gamma f^{\gamma,L}_t} \frac{\sum\limits_{(v')^i} f^{\gamma,L'}_{(v')^i}}{1 - (t+1)^\gamma f^{\gamma,L'}_t}.$$

When $\gamma = 1$ this is

$$P(x^Y, x^{Y'} \leftarrow same|x^Y, x^{Y'} \not\leftarrow X^{Y_{1:t}}) = \sum_i \frac{n^i_vn^i_{v'}}{(L-t)(L'-t)} = \frac{\mathbf{n}^T_v\mathbf{n}_{v'}}{(L-t)(L'-t)}$$

i.e. when $\gamma = 1$ the probability that $x^Y$ came from the same symbol, given that it was overwritten in the second part of both sequence constructions, is just the normalized dot product of the symbol-count vectors for the second parts of the two sequences.

Thus

$$P(x^Y = x^{Y'}|x^Y, x^{Y'} \not\leftarrow X^{Y_{1:t}}) = $$

$$\sum_i \frac{\sum\limits_{v^i} f^{\gamma,L}_{v^i}}{1 - (t+1)^\gamma f^{\gamma,L}_t} \frac{\sum\limits_{(v')^i} f^{\gamma,L'}_{(v')^i}}{1 - (t+1)^\gamma f^{\gamma,L'}_t} + \left(1 - \sum_i \frac{\sum\limits_{v^i} f^{\gamma,L}_{v^i}}{1 - (t+1)^\gamma f^{\gamma,L}_t} \frac{\sum\limits_{(v')^i} f^{\gamma,L'}_{(v')^i}}{1 - (t+1)^\gamma f^{\gamma,L'}_t}\right)\frac{1}{Z}$$

When $\gamma = 1$

$$P(x^Y = x^{Y'}|x^Y, x^{Y'} \not\leftarrow X^{Y_{1:t}}) = \frac{\mathbf{n}^T_v\mathbf{n}_{v'}}{(L-t)(L'-t)} + \left(1 - \frac{\mathbf{n}^T_v\mathbf{n}_{v'}}{(L-t)(L'-t)} \right)\frac{1}{Z}.$$

Thus

$$P(\textrm{Case 2}) = [1 - (t+1)^\gamma f^{\gamma,L}_t][1 - (t+1)^\gamma f^{\gamma,L'}_t] \times$$

$$\left[
\sum_i \frac{\sum\limits_{v^i} f^{\gamma,L}_{v^i}}{1 - (t+1)^\gamma f^{\gamma,L}_t} \frac{\sum\limits_{(v')^i} f^{\gamma,L'}_{(v')^i}}{1 - (t+1)^\gamma f^{\gamma,L'}_t} + \left(1 - \sum_i \frac{\sum\limits_{v^i} f^{\gamma,L}_{v^i}}{1 - (t+1)^\gamma f^{\gamma,L}_t} \frac{\sum\limits_{(v')^i} f^{\gamma,L'}_{(v')^i}}{1 - (t+1)^\gamma f^{\gamma,L'}_t}\right)\frac{1}{Z}
\right]$$

When $\gamma = 1$:

$$P(\textrm{Case 2}) = \frac{L-t}{L+1}\frac{L'-t}{L'+1}\left[\frac{\mathbf{n}^T_v\mathbf{n}_{v'}}{(L-t)(L'-t)} + \left(1 - \frac{\mathbf{n}^T_v\mathbf{n}_{v'}}{(L-t)(L'-t)} \right)\frac{1}{Z}\right].$$

\textbf{Case 3}: $(x^Y \leftarrow Y_{1:t}) \land (x^{Y'} \not\leftarrow Y'_{1:t}) \land (x^Y = x^{Y'})$, i.e. $x$ is untouched in the first sequence but overwritten in the second, but happens to be overwritten by $x^{Y_{1:t}}$. Once again the first two events are independent, so 

$$P(\textrm{Case 3}) = (t+1)^\gamma f^{\gamma,L}_t[1-(t+1)^\gamma f^{\gamma,L'}_t]P(x^Y = x^{Y'}|x^Y \leftarrow Y_{1:t}, x^{Y'} \not\leftarrow Y'_{1:t}).$$

When $\gamma = 1$:

$$P(\textrm{Case 3}) = \frac{t+1}{L+1}\frac{L'-t}{L'+1}P(x^Y = x^{Y'}|x^Y \leftarrow Y_{1:t}, x^{Y'} \not\leftarrow Y'_{1:t}).$$

By similar reasoning as in Case 2, the event $(x^Y = x^{Y'}|x^Y \leftarrow X^{Y_{1:t}}, x^{Y'} \not\leftarrow X^{Y'_{1:t}})$ can occur if (1) $x^Y$ and $x^{Y'}$ are taken from the same symbol $i$ or (2) $x^Y$ and $x^{Y'}$ are taken from different symbols but end up with the same component state.

Following similar logic as in Case 2, except swapping $Y^{t+1:L}$ with $Y^{1:t}$ we thus have

$$P(x^Y = x^{Y'}|x^Y \leftarrow Y_{1:t}, x^{Y'} \not\leftarrow Y'_{1:t}) =$$

$$\left[
\sum_i \frac{\sum\limits_{u^i} f_{u^i}^{\gamma, t} \sum\limits_{(v')^i} f_{(v')^i}^{\gamma,L'}}{1 - (t+1)^\gamma f^{\gamma,L'}_t} 
+ \left(1 - \sum_i \frac{\sum\limits_{u^i} f_{u^i}^{\gamma, t} \sum\limits_{(v')^i} f_{(v')^i}^{\gamma,L'}}{1 - (t+1)^\gamma f^{\gamma,L'}_t}\right)\frac{1}{Z}
\right]$$

When $\gamma = 1$:

$$P(x^Y = x^{Y'}|x^Y \leftarrow Y_{1:t}, x^{Y'} \not\leftarrow Y'_{1:t}) =
\frac{\mathbf{n}^T_{u}\mathbf{n}_{v'}}{(t+1)(L'-t)} + \left(1 - \frac{\mathbf{n}^T_{u}\mathbf{n}_{v'}}{(t+1)(L'-t)} \right)\frac{1}{Z}$$

Thus,

$$P(\textrm{Case 3}) = $$
$$(t+1)^\gamma f^{\gamma,L}_t[1-(t+1)^\gamma f^{\gamma,L'}_t] \times 
\left[
\sum_i \frac{\sum\limits_{u^i} f_{u^i}^{\gamma, t} \sum\limits_{(v')^i} f_{(v')^i}^{\gamma,L'}}{1 - (t+1)^\gamma f^{\gamma,L'}_t} 
+ \left(1 - \sum_i \frac{\sum\limits_{u^i} f_{u^i}^{\gamma, t} \sum\limits_{(v')^i} f_{(v')^i}^{\gamma,L'}}{1 - (t+1)^\gamma f^{\gamma,L'}_t}\right)\frac{1}{Z}
\right]$$

And when $\gamma = 1$:

$$P(\textrm{Case 3}) = \frac{t+1}{L+1}\frac{L'-t}{L'+1}\left[
\frac{\mathbf{n}^T_{u}\mathbf{n}_{v'}}{(t+1)(L'-t)} + \left(1 - \frac{\mathbf{n}^T_{u}\mathbf{n}_{v'}}{(t+1)(L'-t)} \right)\frac{1}{Z}
\right].$$

\textbf{Case 4}: $(x^Y \not \leftarrow Y_{1:t}) \land (x^{Y'} \leftarrow Y'_{1:t}) \land (x^Y = x^{Y'})$, i.e. $x$ is untouched in the second sequence but overwritten in the first, but happens to be overwritten by $x^{Y'_{1:t}}$. This is symmetric to Case 3, so

$$P(\textrm{Case 4}) = $$
$$[1 - (t+1)^\gamma f^{\gamma,L}_t](t+1)^\gamma f^{\gamma,L'}_t \times \left[
\sum_i \frac{\sum\limits_{v^i} f_{v^i}^{\gamma,L}\sum\limits_{u^i} f_{u^i}^{\gamma, t}}{1 - (t+1)^\gamma f^{\gamma,L}_t} 
+ \left(1 - \sum_i \frac{\sum\limits_{v^i} f_{v^i}^{\gamma,L}\sum\limits_{u^i} f_{u^i}^{\gamma, t}}{1 - (t+1)^\gamma f^{\gamma,L}_t}\right)\frac{1}{Z}
\right]$$

And when $\gamma = 1$:

$$P(\textrm{Case 4}) = \frac{L'-t}{L+1}\frac{t+1}{L'+1}\left[
\frac{\mathbf{n}^T_{v}\mathbf{n}_{u}}{(L-t)(t+1)} + \left(1 - \frac{\mathbf{n}^T_{v}\mathbf{n}_{u}}{(L-t)(t+1)} \right)\frac{1}{Z}
\right].$$

\textbf{Sum over cases}

$$P(x^Y = x^{Y'}) = P(\textrm{Case 1}) + P(\textrm{Case 2}) + P(\textrm{Case 3}) + P(\textrm{Case 4}) = $$

$$
(t+1)^\gamma f^{\gamma,L}_t(t+1)^\gamma f^{\gamma,L'}_t + [1 - (t+1)^\gamma f^{\gamma,L}_t][1 - (t+1)^\gamma f^{\gamma,L'}_t] \times$$

$$\left[
\sum_i \frac{\sum\limits_{v^i} f^{\gamma,L}_{v^i}}{1 - (t+1)^\gamma f^{\gamma,L}_t} \frac{\sum\limits_{(v')^i} f^{\gamma,L'}_{(v')^i}}{1 - (t+1)^\gamma f^{\gamma,L'}_t} + \left(1 - \sum_i \frac{\sum\limits_{v^i} f^{\gamma,L}_{v^i}}{1 - (t+1)^\gamma f^{\gamma,L}_t} \frac{\sum\limits_{(v')^i} f^{\gamma,L'}_{(v')^i}}{1 - (t+1)^\gamma f^{\gamma,L'}_t}\right)\frac{1}{Z}
\right]$$

$$+(t+1)^\gamma f^{\gamma,L}_t[1-(t+1)^\gamma f^{\gamma,L'}_t] \times 
\left[
\sum_i \frac{\sum\limits_{u^i} f_{u^i}^{\gamma, t} \sum\limits_{(v')^i} f_{(v')^i}^{\gamma,L'}}{1 - (t+1)^\gamma f^{\gamma,L'}_t} 
+ \left(1 - \sum_i \frac{\sum\limits_{u^i} f_{u^i}^{\gamma, t} \sum\limits_{(v')^i} f_{(v')^i}^{\gamma,L'}}{1 - (t+1)^\gamma f^{\gamma,L'}_t}\right)\frac{1}{Z}
\right]
$$

$$+[1 - (t+1)^\gamma f^{\gamma,L}_t](t+1)^\gamma f^{\gamma,L'}_t \times \left[
\sum_i \frac{\sum\limits_{v^i} f_{v^i}^{\gamma,L}\sum\limits_{u^i} f_{u^i}^{\gamma, t}}{1 - (t+1)^\gamma f^{\gamma,L}_t} 
+ \left(1 - \sum_i \frac{\sum\limits_{v^i} f_{v^i}^{\gamma,L}\sum\limits_{u^i} f_{u^i}^{\gamma, t}}{1 - (t+1)^\gamma f^{\gamma,L}_t}\right)\frac{1}{Z}
\right]
$$

When $\gamma = 1$ this simplifies to

$$P(x^Y = x^{Y'}) =  \frac{t+1}{L+1}\frac{t+1}{L'+1}$$

$$
+ \frac{L-t}{L+1}\frac{L'-t}{L'+1}\left[\frac{\mathbf{n}^T_v\mathbf{n}_{v'}}{(L-t)(L'-t)} + \left(1 - \frac{\mathbf{n}^T_v\mathbf{n}_{v'}}{(L-t)(L'-t)} \right)\frac{1}{Z}\right]
$$

$$+\frac{t+1}{L+1}\frac{L'-t}{L'+1}\left[
\frac{\mathbf{n}^T_{u}\mathbf{n}_{v'}}{(t+1)(L'-t)}+ \left(1 - \frac{\mathbf{n}^T_{u}\mathbf{n}_{v'}}{(t+1)(L'-t)} \right)\frac{1}{Z}
\right]$$
$$
+\frac{L'-t}{L+1}\frac{t+1}{L'+1}\left[
\frac{\mathbf{n}^T_{v}\mathbf{n}_{u}}{(L-t)(t+1)} + \left(1 - \frac{\mathbf{n}^T_{v}\mathbf{n}_{u}}{(L-t)(t+1)} \right)\frac{1}{Z}
\right].
$$

\section{Correlated mask function construction}

We introduced correlations into the mask function $\mu$ (Figure 2d) as follows. We first created a random, sparse, binary matrix $W$ of size $N \times 2N$ and with density $r$ (0.004 in Fig 2d), and assigned our component states $1, ..., Z$ to evenly spaced angles between $0$ and $2\pi$. The mask value $\mu_j(X, X')$ was then determined by (1) multiplying $W$ by the concatenated $2N$-dimensional vector $[X; X']$, (2) taking the circular mean $\phi$ of the result, and (3) letting $\mu_j(X, X')$ be equal to absolute angular distance between $\phi$ and $\pi$, divided by $\pi$. This led to mask function values uniformly distributed between $0$ and $1$ yet which were determined in a smoother way by the values of $X$ and $X'$ than when $\mu$ was sampled i.i.d. Upon quadrupling $N$ (Figure 2d magenta) we equivalently decreased $r$ by a factor of 4.

\section{Code}

All code used in the preparation of this manuscript was written in the Python programming language and is available at \url{https://github.com/rkp8000/crossing_over}.

\end{appendices}


% \section*{References}

% References follow the acknowledgments. Use unnumbered first-level heading for
% the references. Any choice of citation style is acceptable as long as you are
% consistent. It is permissible to reduce the font size to \verb+small+ (9 point)
% when listing the references. {\bf Remember that you can use more than eight
%   pages as long as the additional pages contain \emph{only} cited references.}
% \medskip

% \small

% [1] Alexander, J.A.\ \& Mozer, M.C.\ (1995) Template-based algorithms for
% connectionist rule extraction. In G.\ Tesauro, D.S.\ Touretzky and T.K.\ Leen
% (eds.), {\it Advances in Neural Information Processing Systems 7},
% pp.\ 609--616. Cambridge, MA: MIT Press.

% [2] Bower, J.M.\ \& Beeman, D.\ (1995) {\it The Book of GENESIS: Exploring
%   Realistic Neural Models with the GEneral NEural SImulation System.}  New York:
% TELOS/Springer--Verlag.

% [3] Hasselmo, M.E., Schnell, E.\ \& Barkai, E.\ (1995) Dynamics of learning and
% recall at excitatory recurrent synapses and cholinergic modulation in rat
% hippocampal region CA3. {\it Journal of Neuroscience} {\bf 15}(7):5249-5262.

\end{document}